\documentclass[draft,gc]{agutex}

 \usepackage{graphicx}
 \setkeys{Gin}{draft=false}
\authorrunninghead{RODA ET AL.}

\titlerunninghead{Numerical simulations of an ocean/continent convergent system.}

\authoraddr{M. Roda,
Universit\`a degli Studi di Milano,
Dipartimento di Scienze della Terra ÒA. DesioÓ, Sezione di Geofisica, Via L. Cicognara 7, 20129 Milano, Italy.
(manuel.roda@unimi.it).}
\authoraddr{A. M. Marotta,
Universit\`a degli Studi di Milano,
Dipartimento di Scienze della Terra ÒA. DesioÓ, Sezione di Geofisica, Via L. Cicognara 7, 20129 Milano, Italy.
Universit\`a degli Studi di Milano, Italy.
(anna.maria.marotta@unimi.it).}
\authoraddr{M. I. Spalla,
Universit\`a degli Studi di Milano,
Dipartimento di Scienze della Terra Ò A. DesioÓ Sezione di Geologia, and C.N.R.-I.D.P.A., Via Mangiagalli, 34, 20133, Milano, Italy.
(iole.spalla@unimi.it).}

\begin{document}

%% ------------------------------------------------------------------------ %%
%
%  TITLE
%
%% ------------------------------------------------------------------------ %%

\title{Numerical simulations of an ocean/continent convergent system: influence of subduction geometry and mantle wedge hydration on crustal
recycling.}
%
% e.g., \title{Terrestrial Ring Current:
% Origin, Formation and Decay $\alpha\beta\Gamma\Delta$}
% You may use \\ to break the title over several lines.

%% ------------------------------------------------------------------------ %%
%
%  AUTHORS AND AFFILIATIONS - 2 methods
%
%% ------------------------------------------------------------------------ %%

% Method 1
% For three or fewer author/affiliation blocks, use \author{} and \affil{}

\author{M. Roda}
\affil{Sezione di Geofisica, Dipartimento di Scienze della Terra ÒArdito DesioÓ,
Universit\`a degli Studi di Milano, Italia.}

\author{A. M. Marotta}
\affil{Sezione di Geofisica, Dipartimento di Scienze della Terra ÒArdito DesioÓ,
Universit\`a degli Studi di Milano, Italia.}

\author{M. I. Spalla}
\affil{Sezione di Geologia, Dipartimento di Scienze della Terra ÒArdito DesioÓ,
Universit\`a degli Studi di Milano, Italia.}

% ---------------
% Method 2
% For more than three author/affiliation blocks,
% use \author{\altaffilmark{}} and \altaffiltext{}
% \altaffilmark will produce footnote;
% matching altaffiltext will appear at bottom of page.
% May use \\ to start a new line.

% \authors{R. C. Bales, \altaffilmark{1}
% E. Mosley-Thompson, \altaffilmark{2}
% R. Williams, \altaffilmark{3}
% and J. R. McConnell\altaffilmark{4}}

% \altaffiltext{1}
% {Department of Hydrology and Water Resources, University of Arizona,
% Tucson, Arizona, USA.}
%
% \altaffiltext{2}{Department of Geography, Ohio State University,
% Columbus, Ohio, USA.}
%
% \altaffiltext{3}{Department of Space Sciences, University of Michigan,
% Ann Arbor, Michigan, USA.}
%
% \altaffiltext{4}{Desert Research Institute, Division of Hydrologic Sciences,
% Reno, Nevada, USA.}

%% ------------------------------------------------------------------------ %%
%
%  ABSTRACT
%
%% ------------------------------------------------------------------------ %%

% >> Do NOT include any \begin...\end commands within
% >> the body of the abstract.

\begin{abstract}

The effects of the hydration mechanism on continental crust recycling are analyzed through a 2D finite element thermo-mechanical model. Oceanic slab
dehydration and consequent mantle wedge hydration are implemented using a dynamic method. Hydration is accomplished by lawsonite and serpentine
breakdown; topography is treated as a free surface. Subduction rates of 1, 3, 5, 7.5 and 10 cm/y, slab angles of $30^o$, $45^o$ and $60^o$ and a
mantle rheology represented by dry dunite and dry olivine flow laws, have been taken into account during successive numerical experiments. Model
predictions pointed out that a direct relationship exists between mantle rheology and the amount of recycled crustal material: the larger the viscosity
contrast between hydrated and dry mantle, the larger the percentage of recycled material into the mantle wedge. Slab dip variation has a moderate
impact on the recycling. Metamorphic evolution of recycled material is influenced by subduction style. $T_{Pmax}$, generally representative of eclogite
facies conditions, is sensitive to changes in slab dip. A direct relationship between subduction rate and exhumation rate results for different slab
dips that does not depend on the used mantle flow law. Thermal regimes predicted by different numerical models are compared to PT paths followed by
continental crustal slices involved in ancient and recent subduction zones, making ablative subduction a suitable pre-collisional mechanism for burial and exhumation of continental crust.
\end{abstract}

%% ------------------------------------------------------------------------ %%
%
%  BEGIN ARTICLE
%
%% ------------------------------------------------------------------------ %%

% The body of the article must start with a \begin{article} command
%
% \end{article} must follow the references section, before the figures
%  and tables.

\begin{article}

%% ------------------------------------------------------------------------ %%
%
%  TEXT
%
%% ------------------------------------------------------------------------ %%

\section{Introduction}
In ocean-continent subduction, crustal material dragged into the subduction zone is composed mainly of ocean crust and sediments, trench sediments and continental slices
belonging to the subducting plate (microcontinent) \citep{Ring2003} or early continental collision \citep{Chemenda1995} or continental slices
 tectonically eroded from the overriding plate (ablative subduction) \citep{Polino1990,Tao1992,Spalla1996}. To explain
 the mechanism causing the exhumation of subducted HP and UHP crustal material (either continental and/or oceanic), several models have been developed
 during the last 20 years. These models can be grouped into six main classes: a) crust-mantle delamination \citep{Chemenda1995}, in which crustal rocks exhumation is caused by a continuous underthrusting of eroded portions of the subducting
 plate; b) slab break-off \citep{Ernst1997}, in which the exhumation of crustal slices is consequent to the rebound
 induced slab-breakoff; c) slab retreat \citep{Ring2003}, in which the continuous underthrusting of continental
 slices, belonging to the subducting plate, drives the extrusion of HP continental fragments and the retreat of the slab; d) roll-back slab
 \citep{Brun2008}, in which buoyancy forces combined to trench retreat due to the roll-back of the slab, triggered by subduction rate decrease and slab dip increase
 after collision, drive the exhumation of HP crustal slices; e) decoupling of two main ductile layers \citep{Yamato2008},
 in which mainly negative buoyancy and/or faulting drive exhumation; f) subduction channel flow \citep{Stockhert2005}, in which the exhumation is
 driven by the upwelling flow developed in a low-viscosity mantle wedge.

 Although all of the cited mechanisms give an explanation for the exhumation of
 crustal continental rocks, only the channel flow model takes into account the recirculation of continental slices dragged at high depth by ablation,
 during active subduction before the continental collision. Actually, models a, b and c imply subduction of continental material belonging to the
 subducting plate during the collision of a microcontinent or a thinned continental margin. In these models, a radical change in the thermal
 regime occurs due to upwelling of the asthenospheric mantle (b, c and d; \citep{Cloos1988_2}) and/or continental collision (a, c and d;
 \citep{Cloos1993,Gerya2008a}), controlling the exhumation paths. Moreover, model e implies a large extension of the accrectionary wedge (up to 50 km)
 composed mainly by sediments, a characteristic that is not present in the broader spectrum of subduction zones where higher thicknesses are less then 20 km \citep{Lallemand1999,Clift2004,Guillot2009}.

Several pressure-temperature paths (PT-paths) and geochronological data suggest that continental crust slices are exhumed, often associated with
serpentinized peridotites, during active oceanic subduction \citep{Spalla1996,Zucali2002,Hattori2003}. Petrological models
\citep[e.g.,][]{Schmidt1998,Ernst2008}, numerical simulations \citep[e.g.,][]{Arcay2005,Gerya2002,Stockhert2005} and
seismic images \citep[e.g.,][]{Rondenay2008} reveal that the dehydration process of the oceanic slab and consequent mantle wedge hydration have a
primary role of developing a forced flow along the subduction zone, driven by viscosity and density contrasts. Furthermore, numerical models of
subduction zones \citep{Gerya2005_03,Meda2010} suggest that a large amount of subducted continental material can be exhumed from the subduction wedge
to subsurface structural levels prior to collision.
In addition, natural PT data, from subducted continental crust, have been compared with model predictions only for a portion of the Alps, in which continental material
underwent eclogitization and subsequent exhumation under very low temperature conditions \citep{Meda2010}.

In order to verify if natural PT data from continental crust re-equilibrated under HP-LT conditions in other envisaged
geodynamic scenario, from other orogenic belts and during different times (from about 400 Ma to about 20 Ma, peak age), can be consequent to the
dynamics developed within a hydrated mantle wedge, we perform a parametric analysis using a 2D finite element thermo-mechanical model of ablative
subduction with mantle wedge hydration. Another goal of this numerical experiment is to explore the effects of changes in the subduction rate, slab dip
and mantle rheology on the P/T ratio characterizing the metamorphic climax, on the serpentinized mantle area and on the amount and rate of crustal
exhumation. We present the results of a set of numerical simulations with varying subduction rates (1, 3, 5, 7.5 and 10 cm/y), slab angles ($30^o$,
$45^o$ and $60^o$) and mantle rheology (serpentine, dry dunite and dry olivine flow laws). Finally, numerical model results are compared to PT paths
obtained from ancient and recent subduction zones with different slab dips and convergence velocities to make the geodynamic model more quantitative.

\section{Numerical modeling}
The physics of the crust-mantle system during subduction is described by the coupled equations for continuity, conservation of momentum and
conservation of energy, expressed in the form:
\begin{equation} \label{eq:continuita}
\nabla á \cdot  \overrightarrow{v}=0
\end{equation}
\begin{equation} \label{eq:momento}
\frac{\partial \tau_{ij}}{\partial x_{j}}=\frac{\partial p}{\partial x_{i}} - \rho \overrightarrow{g}
\end{equation}
\begin{equation} \label{eq:energia}
\rho c \left( \frac{\partial T}{\partial t}+\overrightarrow{v} \cdot \nabla T \right)=-\nabla \cdot \left (-K \nabla T \right) + \rho H
\end{equation}
respectively, where $\rho$ is the density, $ \overrightarrow{v}$ the velocity, $p$ the pressure, $ \overrightarrow{g}$ the gravity acceleration,
$\tau_{ij}$ the deviatoric stress tensor, $c$ the thermal capacity at constant pressure, $T$ the temperature, $K$ the thermal conductivity and $H$ the
heat production rate per unit mass. Equations (\ref {eq:continuita}), (\ref {eq:momento}) and (\ref {eq:energia}) are solved by means of the 2D finite
element code SubMar \citep{Marotta2006,Marotta2007}, here modified to account for mantle wedge hydration mechanism and erosion/sedimentation processes, as described in the next paragraph. Integration of equation (2) is performed using the penalty function formulation. Temporal integration of equation (3) is performed using the upwind Petrov-Galerkin method and with a fixed time step of 0.1 My.

\subsection{Geometry and boundary conditions}
The numerical solution of (1), (2) and (3) is performed in a 2-D domain 1400 km wide and 708 km deep (Figure 1), discretized by an irregular grid, composed by 2808 quadratic triangles and 6000 nodes. The linear size of the elements ranges from 2 km, in the mantle wedge area, to 20 km at the margin of the model. Some results of the resolution tests performed initially are presented in Appendix A.
Materials (Table 1) are compositionally differentiated via the Lagrangian particle technique \citep{Christensen1992}. At the beginning of each simulation, 973080 markers, identified by different indexes, are distributed with a density of 1 marker per 0.25 km$^2$ to define the atmosphere, the sediments, the oceanic crust, the continental crust and the lithospheric mantle. The position of the individual particles during the evolution of the system is calculated by applying the first order (both in time and in space) Runge-Kutta scheme (Appendix B).

To study the effects of mantle rheology variations, a dry dunite and a dry olivine flow law are chosen to represent two different unserpentinized mantle rheologies. A constant viscosity of $10^{19}$ $Pa\cdot s$ is chosen to approximate the serpentinized mantle flow law according to \citet{Honda2003}, \citet{Gerya2005_03} and \citet{Meda2010}.
Velocity boundary conditions correspond to zero velocity ($v_x=0$; $v_y=0$) imposed along all boundaries of the 2D domain, with the exception of the uppermost 80 km of the vertical left side, along which $dv_x$/$dx=0$ and $v_y=0$ are imposed. To simulate plate convergence, we fix the velocity along the upper boundary of the oceanic plate, at 0 km depth, varying from 1 to 10 cm/y for the different simulations. In order to force the starting of subduction, we fix the same velocity up to 80 km depth, at the nodes of the numerical grid distributed along a dip ranging from $30^o$ to $45^o$ and $60^o$.

Thermal boundary conditions correspond to fixed temperatures at the 8 km thick uppermost atmospheric layer (from 8 to 0 km depth) and at the bottom of the model domain, 300 K and 1600 K, respectively, and to zero thermal flux through the vertical sidewalls.
The initial thermal configuration corresponds to a constant temperature of 300 K in the atmosphere, a uniform upper purely conductive thermal boundary layer throughout the lithosphere (from 0 to 80 km depth and from 300 K to 1600 K) and a uniform sub-lithospheric temperature of 1600 K. The isotherm 1600 K represents the base of the lithosphere throughout the evolution of the system.

The assumed model features cause a strong coupling (i.e. high friction) between the upper and lower plates, making our
models representative of the sole erosive margin type, which represent about the 60\% of the global active margins \citep{Clift2004}.

\subsection{Topographic surface}
The topographic surface is treated as a free surface between the crust and the low density, low viscosity layer that simulates the atmosphere (8 km
thick). Erosion and sedimentation processes are simulated by evaluating, at each time step, the size of the accrectionary prism, which is generated by
the combined effects of oceanic plate bending and the ablation of the overriding plate. We simulated the instantaneous erosion of continental crust and the consequent (total or partial) filling of the accrectionary prism by using a substitution technique, in which all the crustal
particles lying above 2 km are replaced with air particles, and an equal number of sediment particles are introduced into the trench region (to
respect global mass conservation), This procedure reproduces the erosion and sedimentation rates variable in time, as function of the system dynamics.

\subsection{Hydration and serpentinization of mantle wedge}
Phase diagrams for $H_2O$-saturated average mantle peridotite show that in the serpentine stability field, hydrous phases have the maximum $H_2O$
contents. The water loss related to the serpentine destabilization is greater than 50 wt\%, and the water content approaches zero consequent to
chlorite destabilization \citep{Schmidt1998}. In this study the progressive mantle hydration is controlled by serpentine stability field calculated
from \citet{Schmidt1998} using the following equations:
\begin{equation} \label{eq:hydration1}
y_{hydr}=-0.0008\cdot T_{elem}^2+1.1873\cdot T_{elem}-418.22
\end{equation}
above 72.5 km depth and
\begin{equation} \label{eq:hydration2}
y_{hydr}=0.0016\cdot T_{elem}^2-1.19828\cdot T_{elem}+320.76
\end{equation}
below 72.5 km depth, where $y_{hydr}$ is the maximum hydration depth and $T_{elem}$ is the average temperature predicted in each element. Outside of the
serpentine stability field a newtonian unserpentinized mantle behavior is assumed, represented by the dry viscosity laws of olivine or dunite.

The upper limit of hydration is fixed at 30 km depth, according to \citet{Schmidt1998} who show that the most important
contribution of dehydration occurs below 30 km depth. They indicate a significant water budget available until a depth of about 150--200 km. Nevertheless some subducted continental rocks, returned to the surface during continental collision, show evidence of coesite-bearing assemblages \citep[e.g.,][]{Groppo2009} and stishovite
\citep{Liu2008} that induce to consider the transport of water efficient until greater depths (up to 250-300 km).
For this reason, we simulate the maximum dehydration depth of the oceanic crust using the following lawsonite stability field:
\begin{equation} \label{eq:dehydration}
y_{dehy}=-\frac{T_{elem}+0.8755}{714.55}
\end{equation}
where $y_{dehy}$ is the maximum dehydration depth, below which the water content in hydrous phases in $H_2O$-saturated MORB basalt is negligible
\citep{Schmidt1998}. We applied this equation until a 300 km depth.
The equations cited above define the hydration field, in which we assume that material exhibits a constant viscosity of $10^{19}$ $Pa\cdot s$
\citep{Arcay2005,Gerya2005_03} and the mantle density decrease to 3000 kg/$m^3$ \citep{Schmidt1998,Juteau1999}.
Since less then $20\%$ of the oceanic floor has been estimated to be affected by fracture zones with serpentinization (at depth less then 5-7 km) and thus,
the amount of serpentinized peridotite within the subducting lithosphere is less then $10\%$ \citep{Juteau1999} we do not introduce a further serpentinized level, although some authors suggest that the subducting lithosphere could be serpentinized \citep[e.g.,][]{Yamato2007}.

\section{Model predictions}
In this section we analyze the effects of dip, subduction rate and mantle rheology changes on the
thermal and dynamic evolution of the system (Table 2). We consider to be exhumed continental crust all particles that have been buried below and
recycled above 40 km depth, which is the starting crustal thickness. Thus, for all simulations, we count the recycled markers to obtain the ratios of
buried/exhumed material (exhumation \%), the hydration area width, the average of the maximum and minimum pressure
values reached by particles during burial and exhumation stages ($P_{max}$, $P_{min}$), the temperature values predicted for $P_{max}$ ($T_{Pmax}$) and
the maximum temperature values reached by the recycled particles during the complete burial/exhumation loops ($T_{max}$). We also considered two
exhumation rates: the maximum exhumation rate (Max. exh. rate), which is calculated from the upward velocity vectors recorded along the mantle wedge,
and the total exhumation rate (Total exh. rate), which is obtained from analysis of the PTt-paths using the ratio between ($Y_{P_{max}}$-$Y_{P_{min}}$) and the time span.

\subsection{$30^o$ dip simulations}
For the dry dunite runs, ablative subduction starts in the early stages and affects both upper and lower continental markers (See an example in Figure
2). This dynamics dominate until the serpentinized mantle wedge area exceeds a critical size (from 1800 to 3250 $km^2$, Figure 3) sufficient to develop a
counter-clockwise flow into the mantle wedge, which drives the exhumation process (Figure 2b). During the mature stages of subduction, the exhumation
of continental markers becomes the most important process within the mantle wedge. The combined effect of ablation and upwelling flow induces the
thermal and mechanical erosion of the overriding plate along a zone extending from 100 to about 200 km from the trench. Consequent lower crust
denudation occurs with a contemporaneous accrection of the exhumed crustal material, trench sediments and crustal and lithospheric mantle rocks into
the subduction zone, from the surface to about 50-80 km depth (Figure 2c and f). In the serpentinized mantle wedge, a continuous counter-clockwise flow
allows the recycled crustal markers to reach shallow levels (Figure 2c). Only part of the recycled material is exhumed near the surface: several
particles can remain in a deep portion of the mantle wedge or can be totally buried (Figure 2c).
Generally, two main vortical flows develop: one near the surface and one within the mantle wedge (Figure 2c and f).

The increase of subduction velocities induces a decrease in the dip of the oceanic slab (Figure 4b). Velocity also influences the mantle wedge
kinematics: for slow subduction (1 cm/y) the less efficient counter-clockwise flow is not sufficient to allow the exhumation of a discrete crustal mass.
On the contrary, the accrectionary prism is well developed, and a large sediment mass is buried. The fastest subduction zones
simulated (10 cm/y) show a well-developed counter-clockwise flow, and a large amount of recycled markers is carried up to the surface (Figure 4b). The
accrectionary prism is not well-developed, and a small mass of sediments is buried to depth. At medium subduction rates (3-5 cm/y), the exhumation
percentage is the highest; this is due to the dominance of the upwelling flow on the burial flow. In contrast, for fast subduction, burial flow dominates over the upwelling flow, preventing a large exhumation percentage.

The thermal patterns are also strongly affected by the subduction velocity, with a moderate increase of thermal erosion for fast subduction.
In general, however, we observe a prevalence of conduction rather then advection in all models, due to the low angle of the subducting plate and
consequent low thermal erosion for all subductions.
The hydration area is also affected by the subduction plate velocity, reaching a maximum value for medium subduction rates (3 and 5 cm/y; Figure 3a,
blue dots).

A statistical analysis of recycled crustal particles reveals that the markers reach a maximum mean pressure of 1.4-1.65 GPa (1.85 GPa maximum peak
pressure) before exhumation, with a moderate dependence on the subduction rate. $T_{Pmax}$ values (Figure 5a) range between $350^o$C and $700^o$C and
are similar to $T_{max}$ values recorded by the markers during burial/exhumation loops. This result reflects the low thermal gradients prediction along
the mantle wedge and is due to the low subduction angle.
The exhumation rate generally increases linearly with the subduction rate. Maximum exhumation rates are about 1/3 of subduction velocity (Figure 5c).
Total exhumation rates are one order of magnitude lower then the maximum exhumation rate. This difference is due to the non-linear trajectories of
particles during the upwelling. Although the exhumation rate increases with the subduction rate, low exhumation percentages are recorded for both slow
and fast subductions. The general trend of exhumation percentage, as a function of the subduction rate, is quite similar to that of the hydration area:
this suggests a direct correlation between the hydration area and the amount of exhumed material.

Upon increasing mantle viscosity, achieved by switching from a dry dunite rheology to dry olivine (Table 1), a
decrease of the slab dip is observed. The thermal and kinematic structure is also influenced, with consequent variations in hydration area. In fact,
the hydration area displays a moderate increase for high subduction rates, with an opposite behavior with respect to dry dunite runs (Figure 3a, green
lines). Under a change in rheology the ratio of buried/exhumed particles increases  (but with the same subduction rate
dependence) because of strong viscosity gradients at the transition between the hydrated and dry mantle (Figure 5e, green dots). $P_{max}$ increases to a mean maximum value of 1.8 GPa (2.25 GPa maximum peak pressure) along with increases in $T_{max}$ and $T_{Pmax}$
(Figure 5f, b and a, green dots). Same as hydration area also exhumation percentage displays an increase for fast subduction (7.5-10 cm/y) (Figure 5e, green dots).

\subsection{$45^o$ dip simulations}
Same as in the $30^o$ dry dunite runs, ablative subduction dominates during the early subduction stages whilst, in the late stages, lithospheric mantle denudation together with lower crust occurs (Figure 4c and d).
In contrast to the previous model, slab velocity seems not to affect variations of the dip of the oceanic plate with depth.
Thermal and mechanical erosion of the overriding plate is weak for slower subductions, and the accrectionary prism is well-developed but lesser than in $30^o$ simulations. The fastest subduction (10 cm/y) simulations show a dominant advective mantle flow, which allows significant thermal and mechanical
erosion on the overriding plate, out to 300 km from the trench, with lower crust and lithospheric mantle denudation (Figure 4d).
In slow subduction, thermal gradients along mantle wedge are low, and the counter-clockwise flow is weak (Figure 4c). In
contrast, in fast subduction zones, thermal gradients along the mantle wedge are very high, and isotherms are very close to the slab surface.

The hydration area linearly decreases with increasing subduction rate
(Figure 3b, blue lines). In contrast to the $30^o$ simulation only one counter-clockwise flow cell develops in the mantle wedge
for the different subduction rates.
It is worth stressing that the hydration areas predicted by $45^o$ model (Figure 5b) are one order of magnitude lower
than those predicted by $30^o$ simulation.

Analysis of recycled crustal particles yields a mean $P_{max}$ of 1.45-1.65 GPa with a maximum values for subduction rate of 5 cm/y. A peak pressure of 1.9 GPa is reached for the 5 cm/y model (Figure 6f, blue full dots). The
$T_{Pmax}$ values range between $400^o$C and $650^o$C and are lower then the $T_{max}$ recorded by the markers during burial/exhumation loops,
especially for medium and fast subduction rates (Figure 6a and b, blue full dots). Moreover, $T_{Pmax}$ and $T_{max}$ values are higher than those
predicted by the $30^o$ runs. This result suggests that the increase of advective mantle flow is related to the increase in subduction rate and dip,
generating a strong thermal gradient in the upper part of the mantle wedge.
Like $P_{max}$, the exhumation percentages are also affected by subduction velocity and the maximum values of exhumation are recorded for medium/fast
subduction rates (3-5-7.5 cm/y, Figure 6e, blue full dots).
Although a high exhumation area is recorded for very slow subduction (1 cm/y; Figure 3b), a small amount of exhumed material is predicted (Figure 6e,
blue dots). As already pointed out for $30^o$ simulations, this result can be due to the lower efficacy of upwelling flow with respect to the burial
flow.
As seen for the $30^o$ dip cases, the change in mantle rheology from dry dunite to dry olivine increases the smoothing of the slab angle during
subduction and we observe more exhumed crustal material and higher mean $P_{max}$ values, which reach 1.8-2.0 GPa for medium velocities. The maximum
$P_{max}$ value reaches about 2.35 GPa for the 3 cm/y model (Figure 6e and f, green full dots).

\subsection{$60^o$ dip simulations}
For dry dunite runs, slab geometry displays an increase of slab dip at about 150-200 km depth, until it reaches a near vertical dip (Figure 4e and f).
In contrast to the $30^o$ and $45^o$ cases, ablative subduction of upper and lower continental crust is less important, due to the smaller coupling
between ocean and continent. Counter-clockwise flow is weak and localized in a narrow area from 60 km to 100 km depth (Figure 4f). Thermal erosion is
strong for medium and fast subduction until 200 km from the trench. The accrectionary prism is already less developed and is absent in fast subduction
zones (Figure 3e and f). Advective mantle flow is quite important from the early stages of subduction and thermal and mechanical erosion dominate over
all stages with a rapid denudation of the lower continental crust and lithospheric mantle of the overriding plate, until 250 km from the trench (Figure3c).
As a consequence, only a small percentage of subducted particles is exhumed to the surface (Figure 7e).

As in the previous simulations, a no-hydrated (white) area, is located in the mantle
wedge, entrapped among the upper plate, the ablated continental crust and the hydrated mantle. Part of it is attributable to the P-T conditions de facto out of
the chosen serpentine stability field. In the remnant part, although the predicted local P-T conditions might be favorable to mantle hydration,
hydration actually does not occur since we have fixed the upper limit of hydration at 30 km depth, in agreement with \citet{Schmidt1998}, who showed
that the most important contribution of dehydration occurs below 30 km depth, as already specified in the numerical modeling section.

High dip models show a relative high exhumed/buried particles ratio for slow subduction models (1-3 cm/y) and the ratio decreases with velocity
increase (Figure 7e, blue full dots). Furthermore, only for slow subduction, a discrete
lithospheric mantle-crust mixing occurs into the mantle wedge (Figure 4e).
Analysis on recycled crustal particles show a low $P_{max}$ values (max 1.8 GPa) and a similar range of $T_{max}$ and $T_{Pmax}$ (from $600^o$C to
$800^o$C), both higher then those in the $30^o$ simulations, according to the high increase of advective flow (Figure 7f, b and a, respectively).
An increase of the mantle viscosity, from dry dunite to dry olivine, induces a general decrease of hydration area for low and medium velocities (Figure
5c).

\section{Natural cases}
Twelve case histories from seven different subduction complexes, variably distributed in space and time, were chosen to compare numerical results and
natural PT estimates. All natural cases referred to ocean-continent or ocean-arc subduction zones and we sorted PT data obtained
for exhumed continental rocks regardless the interpreted provenance of the subducted continental material. This has been
done to verify if the natural paleo-geothermal environments are consistent with those simulated during our numerical experiments, to check if ablative
subduction represents a good alternative mechanism with respect to the different geodynamic interpretations proposed for each case.
Slab angles and velocities of ancient subduction zones were
obtained from more recent and widely accepted interpretations, except for the Paleozoic subductions (Urals and Armorica), which are compared with all
simulations due to the lack of suggestions about the subduction dip and rates (Table 3).

For the Alpine subduction, PT estimates from units belonging to the Western (Sesia-Lanzo Zone),
Central (Languard-Campo Unit) and Eastern (Koralpe-Saualpe Unit) Austroalpine domain have been selected. Lithologic affinities between the
protholithts of Austroalpine and Southalpine rocks (Adria Alpine hinterland) suggest that the provenance of this domain was from the
overriding continental plate of the Alpine subduction  \citep[e.g.,][]{Compagnoni1977}.
Geochronological data show that the eclogitic imprint in the internal complexes of the Western Austroalpine is Late Cretaceous
($65\pm5$ Ma) (\citep{Meda2010}, and refs. therein), while in the external complexes is of Eocenic age (49-40 Ma) \citep{DalPiaz2001}. Dating on eclogitic peak of the
meta-ophiolitic units in Western Alps (Zermatt-Saas) show similar ages (50-38 Ma) \citep{Lapen2003}.
These results and the PT evolutions inferred for different complexes of the Sesia-Lanzo Zone
\citep[e.g.,][]{Pognante1989b,Pognante1991,Meda2010}, often supported by detailed fieldwork
\citep[e.g.,][]{Zucali2002a,Zucali2002,Zucali2004,Babist2006}, suggest that the exhumation of this Austroalpine continental crust occurred when
oceanic subduction was still active. In addition, the age of greenschist retrogradation in the
Austroalpine domain (45-37 Ma in Sesia-Lanzo zone) is within the time span of the eclogitic peak recorded in the meta-ophiolites of
Zermatt-Saas, testifying that exhumation occurred under LP/LT conditions during active oceanic subduction \citep[e.g.,][]{Zanoni2008}.
On the base of radiometric data the HP/LT metamorphism recorded in Central and Eastern Austroalpine domains is interpreted as early-Alpine (Late Cretaceous in age, e.g.
Languard-Campo Unit \citep{Gazzola2000} and Koralpe-Saualpe Unit \citep{Thoni1996}).
In addition the thermal regime characterizing the earlier Alpine deformation and the mineral ages in the
Languard-Campo Unit are consistent with an early subduction, preceding the lower T/depth ratio characterizing the majority of early
Alpine eclogites \citep{Gazzola2000}.
For these reasons, the Austroalpine rocks, affected by HP Alpine metamorphism, have been mainly interpreted as the result of an ablative subduction
process active at the margin of the overriding Adria plate during pre-collisional stages of the Alpine convergence.
This is in agreement with geodynamic models at the scale of the whole belt, implying subduction and exhumation of the
Austroalpine crust before continental collision \citep{Platt1987,Polino1990,Spalla1996,Stockhert2005}.

For the Aegean subduction, continental rocks of the Cyclades Blueschists Unit and of the Phyllite-Quartzite Unit are taken into account. The Cyclades
Blueschists Unit consists of metasediment originally belonging to the Cycladic passive margin, which was the eastern part of the Adria basement with
an HP/LT metamorphic imprint (55-60 Ma in age) \citep{Ring2003}. The Phyllite-Quartzite Unit is a Carboniferous to Middle Triassic sedimentary sequence
with an HP/LT metamorphic imprint (19-24 Ma in age). This unit represents a former part of the Adria micro-continent \citep{Thomson1998}. The
progressive rejuvenation of HP/LT toward South, the southern arc migration and the intermingling of both oceanic and continental exhumed slices have
been justified using slab retreat \citep{Ring2003} or rollback slab \citep{Brun2008} models.
These mechanisms are assumed to explain also the metamorphic evolution of the Turkey units, which have been involved in Aegean subduction.

For the Calabrian examples, HP continental metasediments and marble have been interpreted as the exhumation of HP rocks along an orogenic wedge by
\citet{Rossetti2004}, while \citet{Ring2003} suggest a rollback slab mechanism to explain the exhumation of the HP/LT Calabrian units and related
back-arc extension of the southern Tyrrhenian area.
For the Alaskan Brooks Range Metamorphic Belt, an arc-continent collision is proposed to justify the exhumation of HP/LT continental rocks
\citep{Till1996}.
For the metapelites of the Ile de Groix belonging to the Armorica Massif, we take into account the possibility that they could derive from continental sediments \citep{Bernard1986}. Thrusting of HP units (Upper Unit) onto low-grade units (Lower Unit) is proposed to explain the exhumation of these rocks occurring after the HP event, implying bulk horizontal shortening \citep{Bosse2002}.
\citet{Leech2000} suggest a synconvergent exhumation of the HP and UHP metasediment belonging to the East European platform, accomplished by a
combination of west-directed thrusting followed by normal faulting.

The characteristics of these natural subduction zones are summarized in Table 3; rock types, slab dips, convergent velocities, P-T climax, exhumation
rates and relative references are reported. To infer the slab dip of the ancient Calabrian and Alpine subduction zones,
the west-east upper mantle flow criterion \citep{Doglioni2007} is used. As was already pointed out, Paleozoic subductions have been compared with the
predictions obtained from all the simulations.

Simulated PT climax assemblages of slow subduction rates and low to moderate subduction angles (Aegean, Alps and Turkey) are in good agreement with
numerical predictions, especially for a 3 cm/y slab velocity. Temperatures, in turn, are in the simulation range, regardless of the subduction rate.
Very HP cases, like the Aegean Cyclades Blueschists and Alps1, are in agreement with both $30^o$ and $45^o$ slab dip models for a velocity of 3 cm/y.
Steeper dips seem to be a better configuration for the Calabrian subduction. The metamorphic evolution of the Alaskan and the Paleozoic subduction
zones show a good agreement with PT predictions of moderate to fast subduction simulations, without a particular dip constraint
(Figure 8).

As a function of subduction velocity, natural exhumation rates are comparable to the simulated exhumation rates, as obtained from PTt-paths (total
exhumation rates, Figure 9).
Only Aegean1 shows an exhumation rate higher than those predicted, although it is of the same order of magnitude. The
maximum exhumation rates, which are obtained from the modulus of the upwelling flow vectors, are one order of magnitude higher then those obtained
from PTt-paths, and they are not compatible with the natural exhumation rates (Figure 9). This discrepancy could be
due to the non-linear trajectories of crustal particles during the exhumation stages.

We find generally good agreement between natural PT evolution and numerical predictions. In addition, the Paleozoic systems are characterized by PT
estimates compatible with predictions from fast subduction zones ($>$5 cm/y). Natural exhumation rates are comparable to the predicted ones, if
obtained from the analysis of simulated PTt-paths. Moreover, exhumation rates are one order of magnitude lower then the subduction rates. This finding
is in contrast with some natural examples of UHP, like Dora-Maira, in the Western Alps \citep{Rubatto2001}, Papua New Guinea \citep{Baldwin2004}  and
Tinos, in the Aegean \citep{Brun2008}, which all show exhumation rates equal to or higher than the subduction rates. Also, some numerical models
\citep{Gerya2008a,Warren2008} display high exhumation rates. Nevertheless, the evolution of both Dora-Maira and Papua
New Guinea and the cited simulations refer to collisions, when the amount of crustal mass buried at depth might generate a transient channel into the
mantle wedge with high temperature and strong positive buoyancy, allowing for fast exhumation of the crustal material \citep{Gerya2008a}. Alternatively,
\citet{Brun2008} indicate that high exhumation rates (of up to 30 mm/y) characterize the first stage of the exhumation of UHP rocks. This finding is
in agreement with the results of the numerical model of \citet{Yamato2008} which shows a high exhumation rate (about 1 cm/y) only for the first stage
of the exhumation path. As reported by \citet{Zucali2002}, the exhumation rates obtained from complete PTt-paths of early ocean/continent subduction
stages are one order of magnitude lower than subduction rates, as pointed out by the worldwide subduction review by \citet{Agard2009}. Thus, the higher
exhumation rates may refer to either collisional events or to pre-collisional transient stages.

\section{Comparison with other numerical models of subduction zones}
In this section we compare our results with the results of other numerical models accounting for pre/early-collisional
exhumation of crustal rocks. Specifically, we focus on subduction flow models \citep[e.g.,][]{Stockhert2005, Faccenda2009} and on exhumation in
accrectionary wedge models \citep[e.g.,][]{Yamato2007}.

Pre-collisional exhumation of HP-LT crustal rocks is treated in \citet{Gerya2005_03} and \citet{Stockhert2005} using a
2D finite difference model in which mantle wedge hydration is included. Our model takes into account similar dehydration/hydration mechanisms and
comparable rheological parameter for the mantle wedge material. However we implemented a different numerical approach using a finite element method
with Lagrangian particles technique instead of finite differences and marker-in-cell particle technique. Furthermore we forced the subduction using
fixed velocities until 80 km depth upon the oceanic crust in contrast to a fixed weak zone imposed by \citet{Gerya2005_03}. Moreover we have not
implemented a fixed erosion/sedimentation rates but they are variable within the subduction dynamics. In order to contribute to fill the gap about the
influence of subduction setting on crustal exhumation we performed a parametric analysis on variable dip, velocity and mantle rheology.

In agreement with \citet{Stockhert2005} and \citet{Faccenda2009} we obtained pre-collisional exhumation of continental
crust belonging to the overriding plate dragged to high depth by ablative subduction and also reproduced marble-cake configuration of the accrectionary
and mantle wedges.
In contrast, we obtained lower exhumation rates, more comparable with those obtained by natural data and in addition
we evaluate the fitting between natural PT data and the model predictions on geothermal gradient. Same as shown by \citet{Gerya2005_03} we found a high
efficiency of ablative subduction process to drag large amount on continental material to deep levels. We simulated a strong coupled plates subduction
zone in order to represent an erosive margin \citep{Clift2004} and our results suggest that the amount of ablated material is proportional to the area
of the plates contact and thus to the subduction dip. Furthermore, the predicted efficiency of the ablation should be considered as a maximum
end-member for the erosive margins. Moreover, \citet{Gerya2008}, \citet{Faccenda2008} and \citet{Faccenda2009} pointed out that the introduction of
shear heating and water percolation of hydrated sediments, located into the accrectionary prism, may reduce the friction between the facing plates, up to
totally decoupling. Thus, ablation decreases and a one-side subduction, characterized by an accrectionary margin, develops \citep{Clift2004}.

Within the accretionary wedge models, \citet{Yamato2007,Yamato2008} pointed out the role played by the accrectionary
prism on the exhumation of sediments and continental rocks. However, both the amount and the $P_{max}$ of the exhumed materials strongly depend on the
existence of an accrectionary prism larger then those at present described in natural environments \citep{Cloos1988,Guillot2009}. In addition the accrectionary prism dimension depends on slab dip, subduction rate, coupling/decoupling and sedimentation/erosion rates on the size of the accrectionary prism \citep{Faccenda2009}. Furthermore,
differently from our study, no continental crust exhumation occurs before the collision.

\section{Summary and conclusions}
The parametric analysis performed for ocean/continent convergent systems points out that ablative subduction is a highly efficient process for dragging large masses of upper and lower continental crust, belonging to the overriding plate, to deep levels.
In particular, ablative subduction is much more efficient for low slab dip due to the stronger coupling between the oceanic and continental plates. The
subduction rate has also been observed to have a moderate influence on ablative subduction efficiency.
In addition to slab dip, in $30^o$ simulations, the similarity of trends in hydration area and exhumation percentage
with respect to subduction rates suggests that the width of the hydration area is correlated to the amount of exhumed material.
Maximum values of exhumation percentage are predicted for intermediate subduction rates except in the $60^o$ simulations, where the peak values are
predicted for the lowest velocities (1 cm/y).

The maximum values of $P_{max}$ are recorded by markers of $45^o$ subductions, and the lowest values are predicted by $60^o$ subduction markers. Just
as for the exhumation percentage values, the maximum $P_{max}$ is also generally predicted to occur for medium subduction rate (3-5 cm/y) in low and
intermediate dip simulations ($30^o$ and $45^o$).
$T_{Pmax}$  and $T_{max}$ values display a positive correlation with the subduction angle and a low variation with subduction rate. In $30^o$
simulations thermal gradients along the mantle wedge are low and $T_{Pmax}$ and $T_{max}$ values are comparable within the same subduction rates. In
contrast, in $60^o$ simulations thermal gradients increase along the mantle wedge, and strong thermal and mechanical erosion occur at the base of the overriding plate.
The accrectionary prism is strongly affected by the subduction rate and the subduction dip: increasing both parameters decreases the width of the prism and is accompanied by an increase in trench depth.
Variations in mantle rheology affect the slab geometry for low and moderate subduction dips. We also predict a strong effect on $P_{max}$ as recorded
by crustal markers. Mantle viscosity has a positive correlation with $P_{max}$ values due to the strong viscosity gradient generated at the transition
between hydrated and dry mantle.

It is worth stressing that the small scale circulation in the wedge area, allowing a huge amount of subducted crustal
markers to come back at shallow depths, occurs because of the low-viscosity subduction channel induced by the mantle hydration. The crucial role played by fluids in the whole process has been already investigated in previous works \citep[e.g.,][]{Gerya2002,Meda2010}; in particular, \citet{Meda2010}
show that no exhumation is possible without mantle hydration, before continental collision.

Natural data and model predictions are in good agreement: the thermal states simulated for ablative subduction with a hydrated mantle wedge justify the
natural PT estimates obtained on continental crust units involved in ocean/continent subduction systems. Similarly, the exhumation rates obtained from
analysis of PTt-paths are more compatible with natural ones than those obtained from the upwelling flow vector which could justify only a transient
exhumation stage. In general, numerical simulations and natural data show exhumation rates lower then subduction rates.

The good agreement obtained between our model and models developed using different numerical and
starting approaches \citep[e.g.,][]{Gerya2005_03,Stockhert2005,Faccenda2008,Faccenda2009}, except for the exhumation rates, indicates the robustness of the corner-like flow mechanism on pre-collisional exhumation of the continental crust.
On the basis of these results, we propose ablative subduction of the overriding continental plate with corner-like flow as a good alternative,
pre-collisional mechanism for the subduction and exhumation of continental crust, in contrast to collisional processes such as slab retreat, rollback
slab or microcontinent collision.

\section*{Appendix A: Resolution Tests}
In this section we present the results of the resolution test performed at the beginning of our numerical analysis to verify the
persisting robustness of the used numerical code, after the introduction, within the original SubMar code \citep{Marotta2006},
of the mantle hydration mechanism, that drives small scale wedge circulation and exhumation of subducted crustal material.
We compare the results, in terms of velocity, temperature and stream line and some statistical parameters ($T_{Pmax}$,$T_{max}$, max. exh. rate,
$P_{max}$), obtained by three different grids, with different resolution in the wedge area.
Figure 10, panels $a_1$, $a_2$ and $a_3$, compares the thermo-dynamic settings obtained in the wedge area by the grid used in
the present study (grid 2, panel $a_2$) with those obtained by a rougher grid (grid 1, panel $a_1$, minimum linear size = 5 km)
and a refined grid (grid 3, panel $a_3$, minimum linear size = 1 km). At large scale, both the temperature and velocity fields
appear unaffected by the grid size. Furthermore, focusing on the wedge area, it is possible to enlighten a stable circulation,
although much more defined when grid 2 and grid 3 are used (panel $a_1$, $a_2$, $a_3$, black stream lines). In this case, a huger
amount of crustal material is recycled. For what concerns the statistical parameters, panels $b_i$ generally indicate a high
variability of predicted values of parameters from grid 1 to grid 2, while low changes occurs when results by grid 3 are compared
with those obtained by grid 2, always within the standard deviation of the model.
For this reason and for the more convenient computational time, we chose grid 2 for developing our numerical analysis.

\section*{Appendix B: Compositional differentiation}
The position of the individual particles during the evolution of the system is calculated by applying the first order (both in time and space) Runge-Kutta scheme to equation $d\vec{x}/dt=\vec{v}$, with $\vec{x}$ and $\vec{v}$ indicating the instantaneous position and velocity of each particle, the last one evaluated by interpolating the velocities at the 6 nodes delimiting the element containing the marker through the same shape functions used in the numerical approximation. At each time step, a non-dimensional function $C$ describes the elemental composition, calculated on the basis of the number of different type particles present in the element, such that
\begin{equation}
C=1-C_o-C_c-C_s-C_a
\end{equation}
with $C_i=M_i/M_0$, where $M_i$ is the number of markers of type i (mantle i=m, continental crust i=c, oceanic crust i=o, sediments i=s and atmosphere i=a) within the element and $M_0$ is the maximum number of markers (of any type) that the element can contain.

By taking into account the elemental composition, the density of each element is calculated as
\begin{equation}
\rho=\rho_0^m\left[1-\alpha\left(T-T_0\right)\right]+\nabla\rho_o^mC_o+\nabla\rho_c^mC_c+\nabla\rho_s^mC_s+\nabla\rho_a^mC_a
\end{equation}
below the initial zero topographic level, or
\begin{equation}
\rho=\rho_0^a+\nabla\rho_o^aC_o+\nabla\rho_c^aC_c+\nabla\rho_s^aC_s+\nabla\rho_0^mC
\end{equation}
above the initial zero topographic level, where $\rho_0^m$ and $\rho_0^a$ are the reference densities of mantle and atmosphere, respectively, $\nabla\rho_i^m=(\rho_i-\rho_m)$ and $\nabla\rho_i^a=(\rho_i-\rho_a)$ represent the difference between the density of the different materials ($\rho_i$) and the reference density of the mantle ($\rho_m$) or the atmosphere ($\rho_a$). $\alpha$ is the thermal expansion coefficient.
For what concerns the rheology, a viscous behavior is assumed for the whole system and the viscosity of each element is calculated as
\begin{equation}
\mu=\mu_m\left[1-C_o-C_c-C_s-C_a\right]+\mu_oC_o+\mu_cC_c+\mu_sC_s+\mu_aC_a
\end{equation}
below the initial zero topographic level, or
\begin{equation}
\mu=\mu_aC_a+\mu_oC_o+\mu_cC_c+\mu_sC_s+\mu^mC
\end{equation}
above the initial zero topographic level, with
\begin{equation}
\mu_i=\mu_i^0 e^{\frac{E_{ai}}{n_iR}\left(\frac{1}{T}-\frac{1}{T_0}\right)}
\end{equation}
where $\mu_i^0$ is the reference viscosity of generic material \emph{i} at the reference temperature $T_0$, $E_a^i$ the activation energy and $T$ the temperature.
Since only a viscous rheology is considered, the crust is expected to remain strong at much higher stresses than the brittle limit and we introduce a viscosity cut-off value of $10^{25}$ Pa$\cdot$s to minimize this effect.

%%% End of body of article:

%%%%%%%%%%%%%%%%%%%%%%%%%%%%%%%%
%% Optional Appendix goes here
%
%%%%%%%%%%%%%%%%%
% Geophysical Research Letters only allows an appendix without a letter.
%% You can get this result with
%  \section*{Appendix}
%  or
%  \section*{Appendix: Title}
%%%%%%%%%%%%%%%%%
%
% \appendix
 %resets counters and redefines section heads
% but doesn't print anything.
% After typing  \appendix
%
 %\section{Lagrangian markers calculation and boundary parameters}
% will print
% Appendix A: Here Is Appendix Title

% \section*{Appendix}
% will print
% Appendix
%
% \section*{Appendix: Here Is Appendix Title}
% will print
% Appendix: Here Is Appendix Title
%
% For only 1 appendix \appendix \section{Appendix} is preferred.
% which will print
% Appendix A

%%%%%%%%%%%%%%%%%%%%%%%%%%%%%%%%%%%%%%%%%%%%%%%%%%%%%%%%%%%%%%%%
%
% Optional Glossary or Notation section, goes here
%
%%%%%%%%%%%%%%
% Glossary only allowed in Reviews of Geophysics
% \section*{Glossary}
% \paragraph{Term}
% Term Definition here
%
%%%%%%%%%%%%%%
% Notation -- End each entry with a period.
% \begin{notation}
% Term & definition.\\
% Second Term & second definition.
% \end{notation}
%%%%%%%%%%%%%%%%%%%%%%%%%%%%%%%%%%%%%%%%%%%%%%%%%%%%%%%%%%%%%%%%
%
%  ACKNOWLEDGMENTS

\begin{acknowledgments}
The constructive criticism of Faccenda M. and of an anonymous reviewer greatly improved the text. Prin 2008 "Tectonic trajectories of subducted lithosphere in the Alpine collisional orogen from structure, metamorphism and litho-stratigraphy" and PUR 2008 "La ricerca geofisica: esplorazione, monitoraggio, elaborazione e modellazione" are gratefully acknowledged.
\end{acknowledgments}

%% ------------------------------------------------------------------------ %%
%
%  REFERENCE LIST AND TEXT CITATIONS
%
% Either type in your references using
% \begin{thebibliography}{}
% \bibitem
% Text
% \end{thebibliography}
%
% Or,
%
% If you use BiBTeX for your References, please produce your .bbl
% file and copy the contents into your paper here.
%
% Follow these steps:
% 1. Run LaTeX on your LaTeX file.
%
% 2. Run BiBTeX on your LaTeX file.
%
% 3. Open the new .bbl file containing the reference list and
%   copy all the contents into your LaTeX file here.
%
% 4. Comment out the old \bibliographystyle and \bibliography commands.

% 5. Run LaTeX on your new file before submitting.
%
% AGU does not want a .bib or a .bbl file, but asks that you
% copy in the contents of your .bbl file here.

%\begin{thebibliography}{}
%\bibliographystyle{agu08}
%\bibliography{biblioroda}
%\end{thebibliography}

\newpage

\end{article}

\begin{figure}
\noindent\includegraphics[width=20pc]{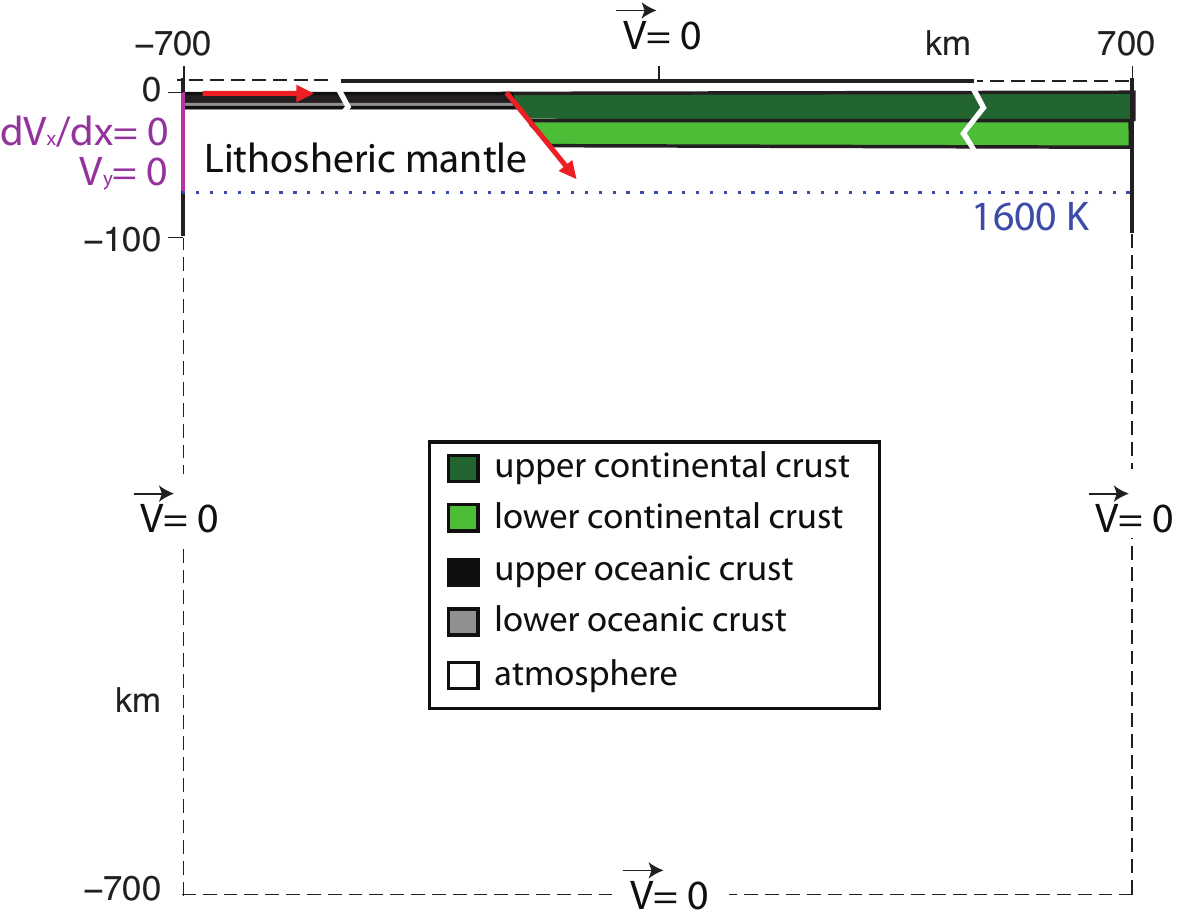}
\caption{Model setup and boundary conditions for all of the simulations. See text for further details. The legend illustrates the marker provenance.}
\end{figure}

\begin{figure}
\noindent\includegraphics[width=32pc]{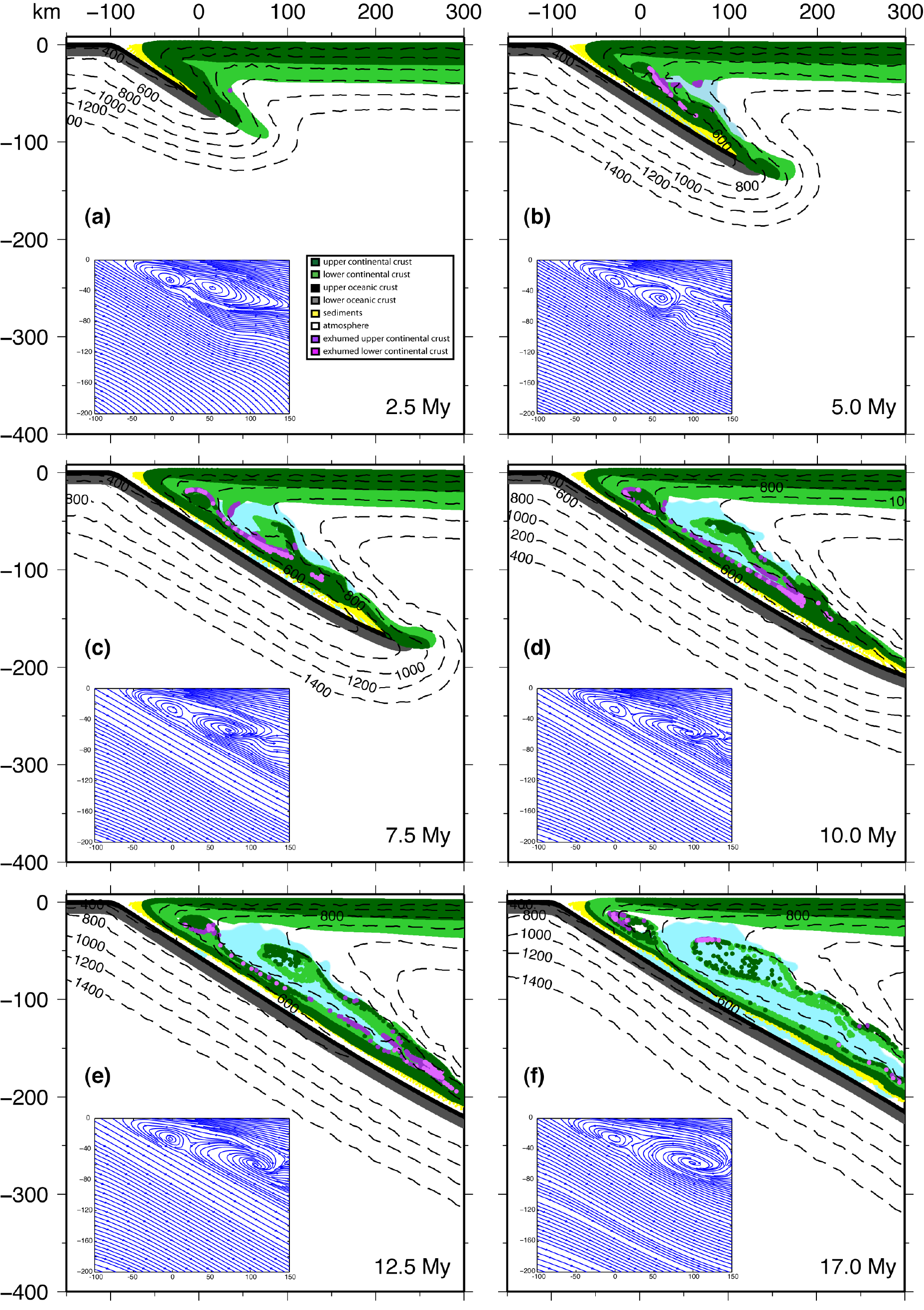}
\caption{Evolution of run 015 (see Table 2) in six different stages (2.5 (a), 5 (b), 7.5 (c), 10 (d), 12.5 (e) and
17.0 (f) Ma). The legend illustrates the marker provenance. In the insets, streamlines (blue lines) obtained for the corner flow are reported.}
\end{figure}

\begin{figure}
\noindent\includegraphics[width=16pc]{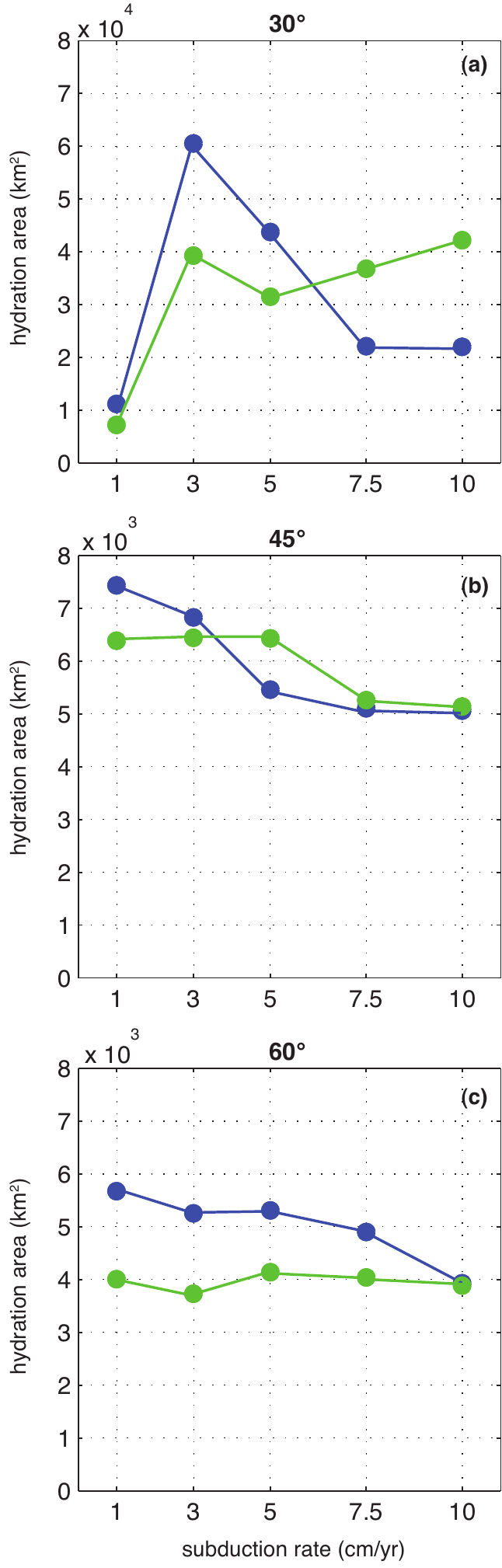}
\caption{Hydration area vs. subduction rate for each slab dip: the blue points refer to dry dunite models and green points to dry olivine models.}
\end{figure}

\begin{figure}
\noindent\includegraphics[width=32pc]{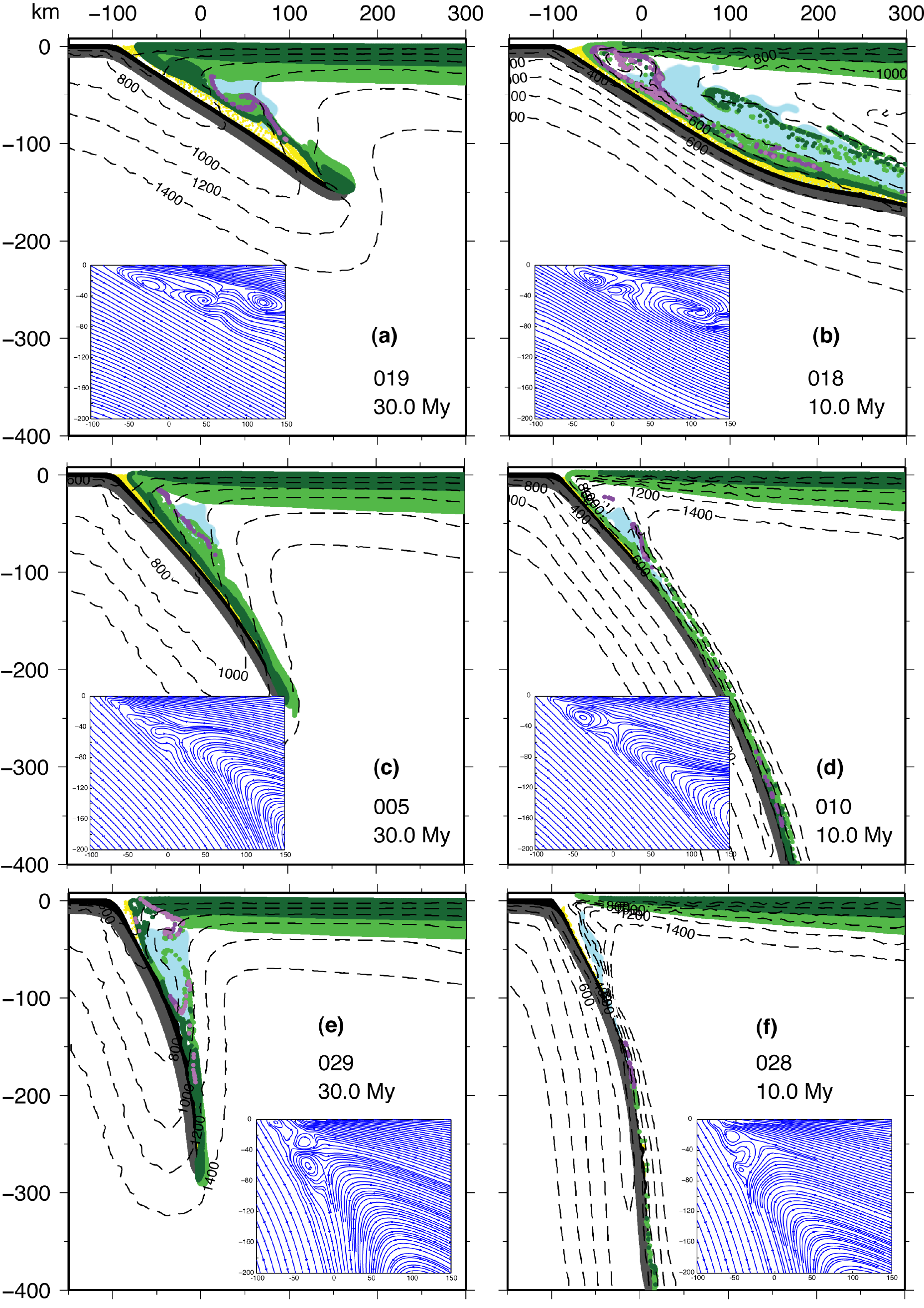}
\caption{Final stage for runs 019, 018 (slab dip $30^o$, a, b), 005, 010 (slab dip $45^o$, c, d) and 029, 028 (slab
dip $60^o$, e, f). Legend as in Figure 2. In the insets, streamlines (blue lines) obtained for the corner flow are reported.}
\end{figure}

\begin{figure}
\noindent\includegraphics[width=32pc]{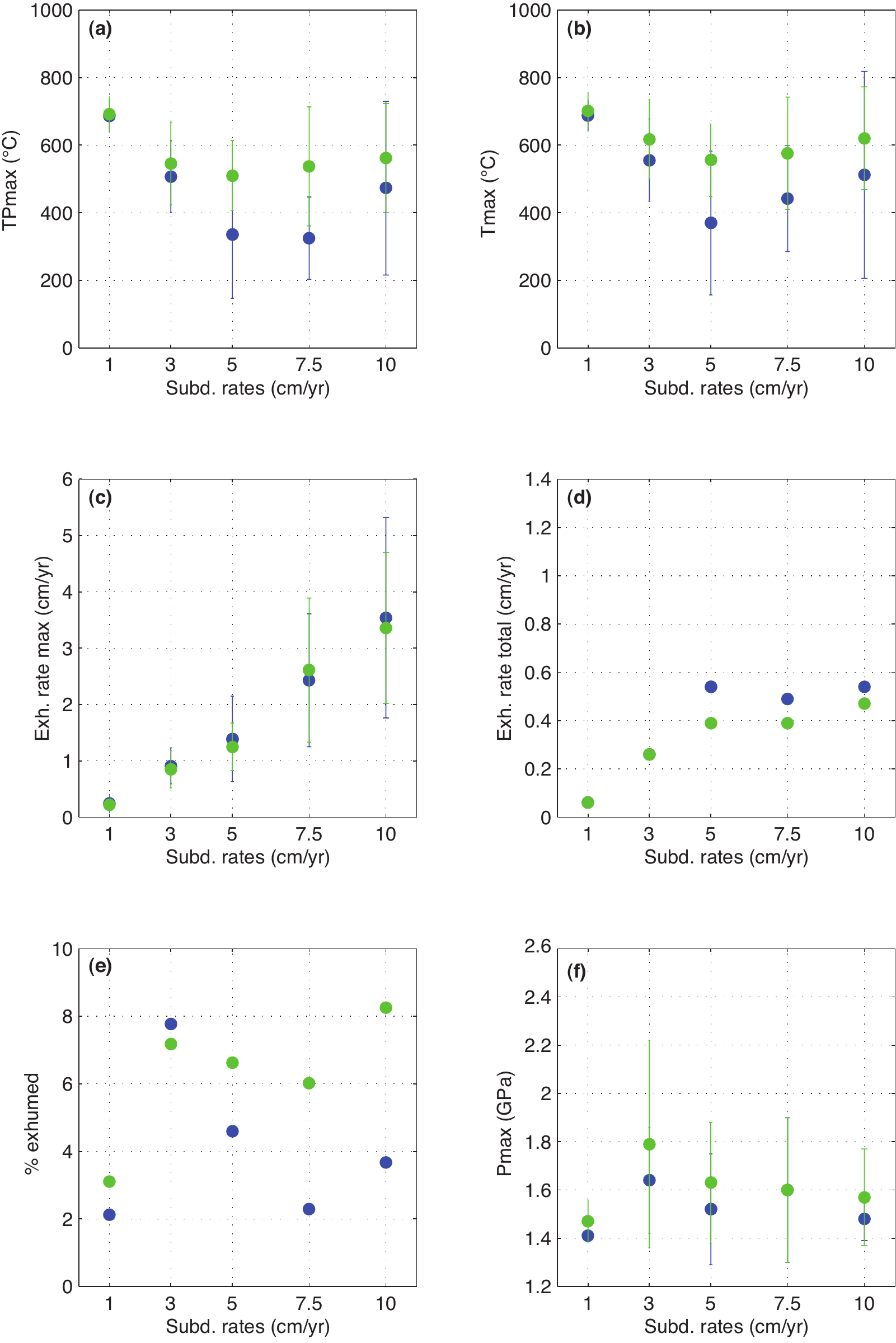}
\caption{Statistical analysis of the exhumed crustal particles for $30^o$ simulations: blue points refer to dry dunite models and green points to dry
olivine models; vertical lines are the error-bars defined as standard deviation of mean values.}
\end{figure}

\begin{figure}
\noindent\includegraphics[width=32pc]{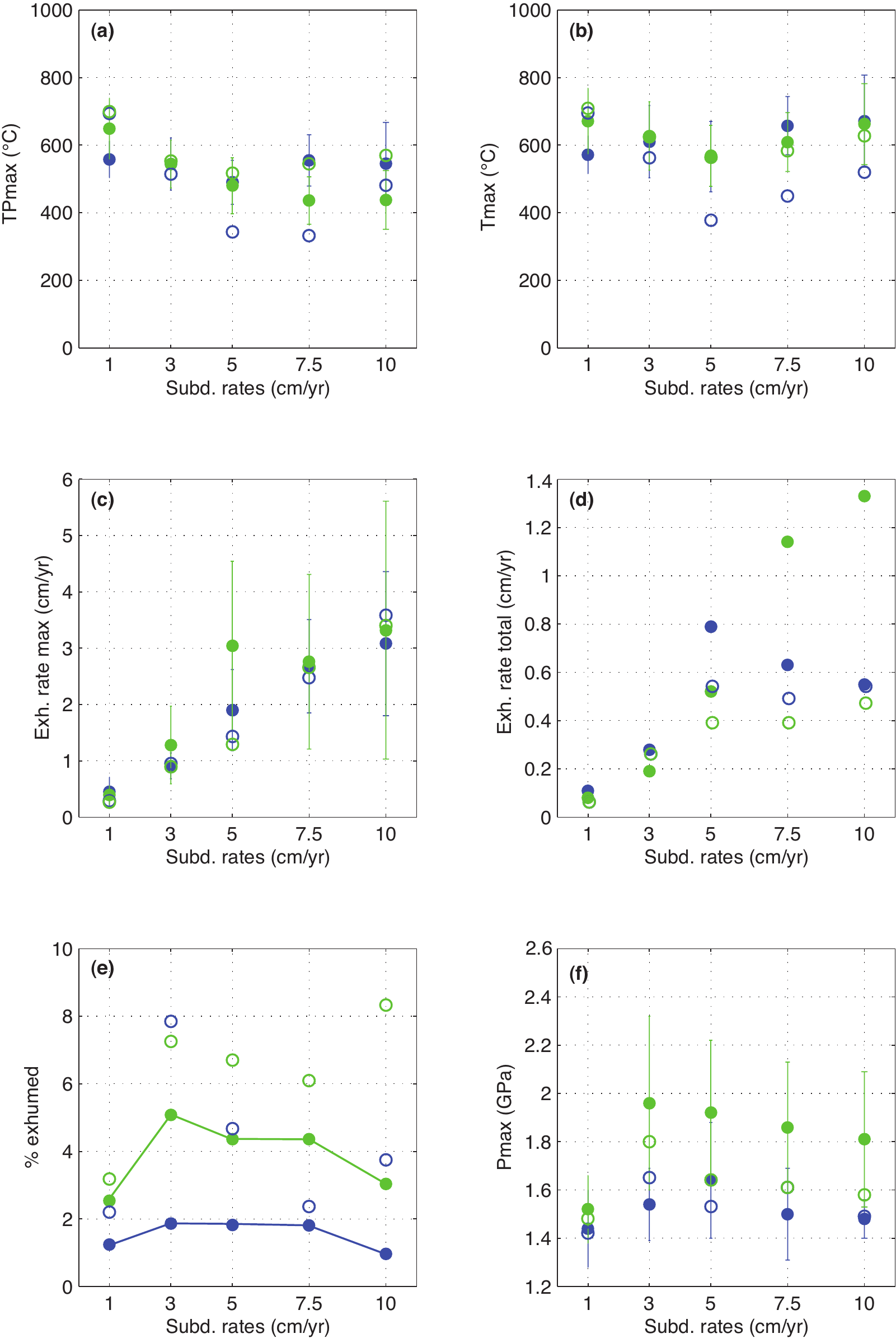}
\caption{Statistical analysis of exhumed crustal particles for $45^o$ simulations: blue points refer to dry dunite models and green points to dry
olivine models; vertical lines are the error-bars defined as standard deviation of mean values. Empty circles are referred to $30^o$ simulation
data.}
\end{figure}

\begin{figure}
\noindent\includegraphics[width=32pc]{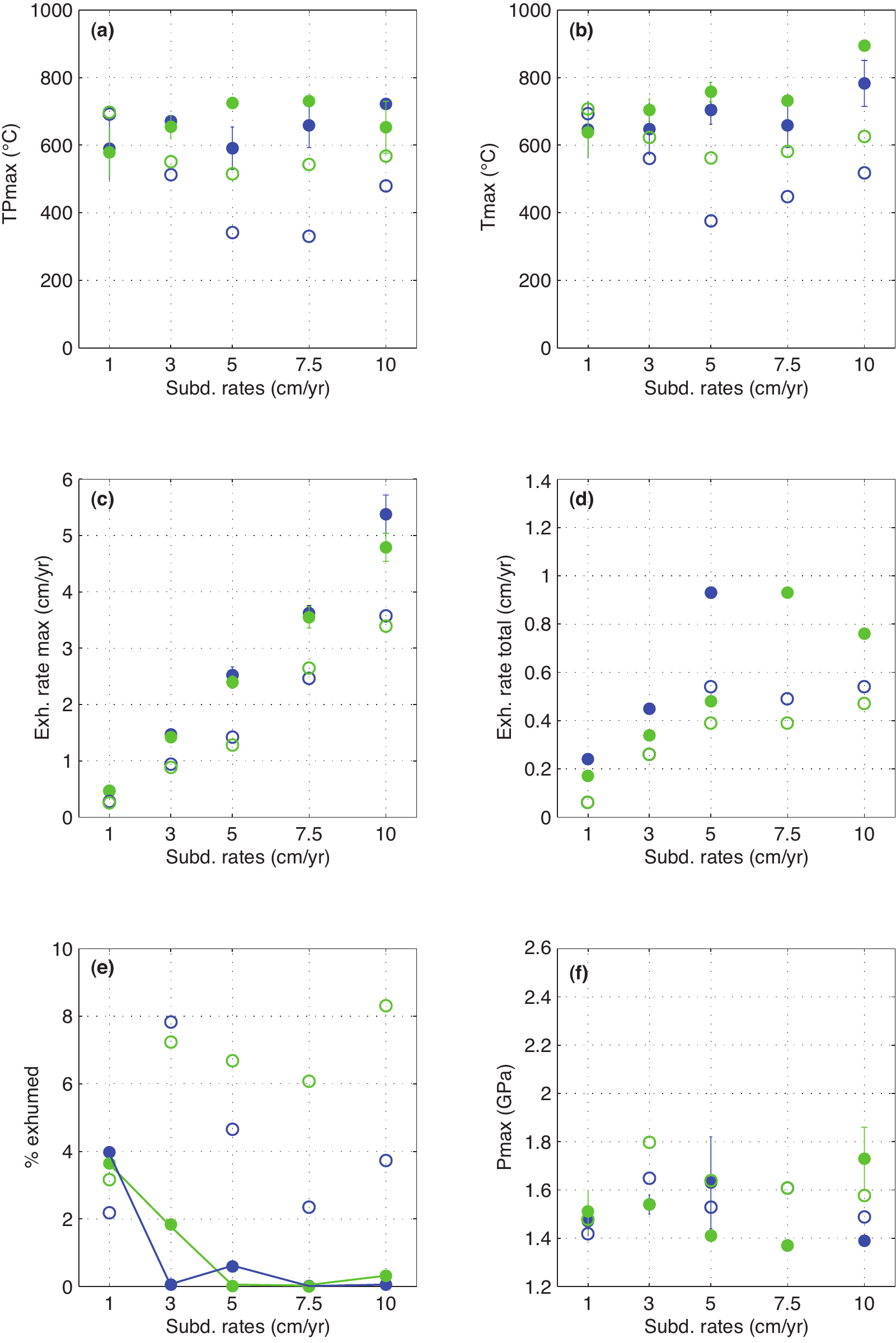}
\caption{Statistical analysis of exhumed crustal particles for $60^o$ simulations: blue points refer to dry dunite models and green points to dry
olivine models; vertical lines are the error-bars defined as standard deviation of mean values. Empty circles are referred to $30^o$ simulation
data.}
\end{figure}

\begin{figure}
%\begin{landscapefigure}
\noindent\includegraphics[width=32pc]{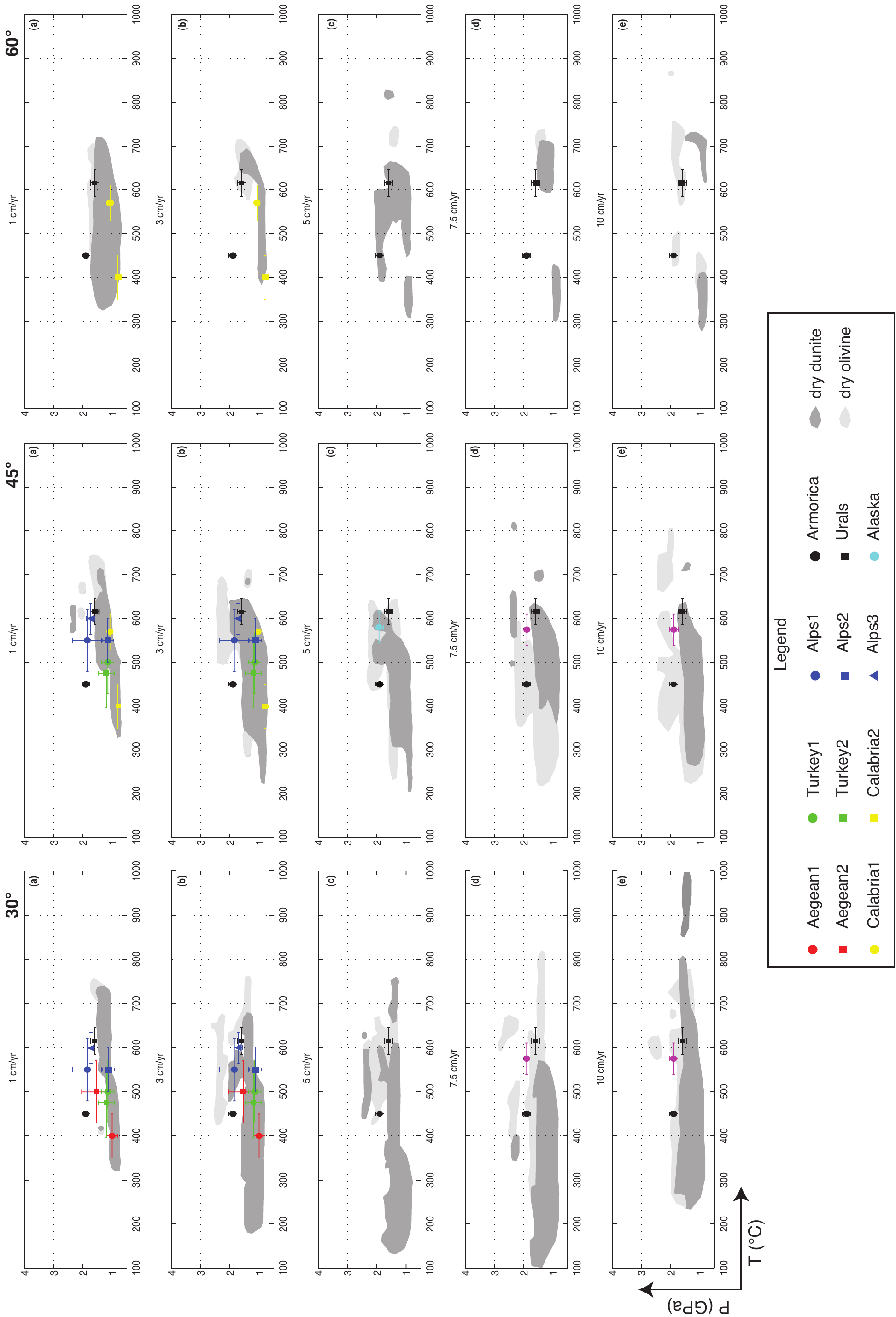}
\caption{Comparison of simulated vs. natural PT estimates: colored symbols refer to units listed in Table 3. Vertical lines are the error-bars according to the original estimates.}
% \end{landscapefigure}
\end{figure}

\begin{figure}
\noindent\includegraphics[width=16pc]{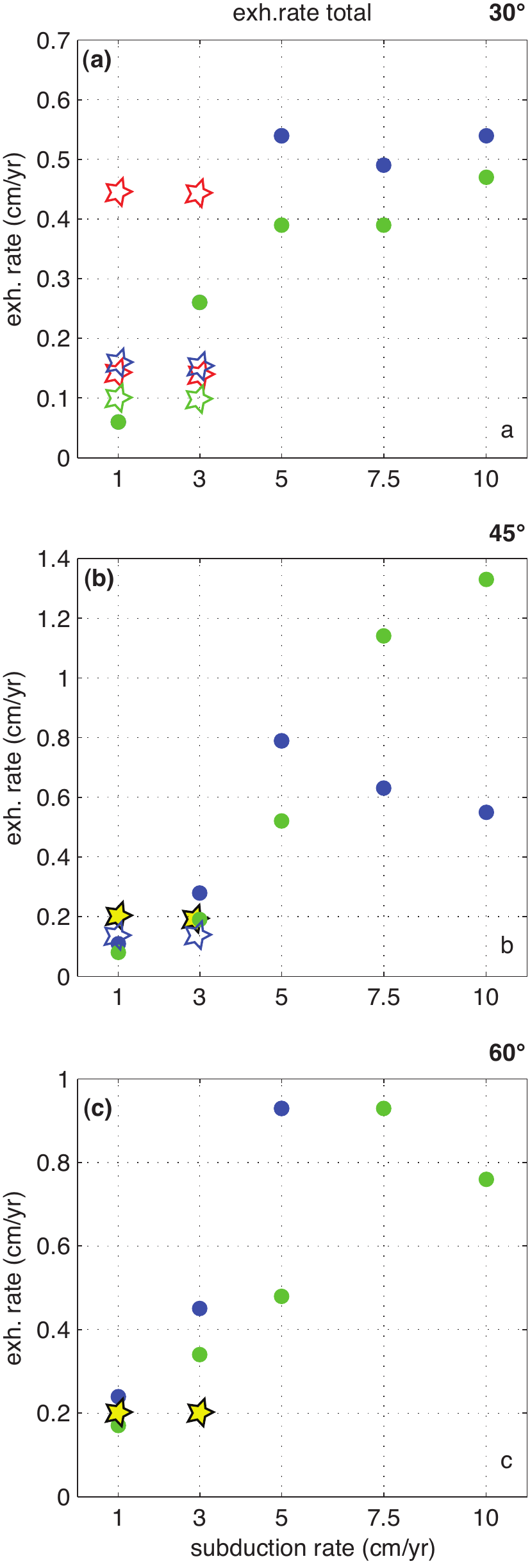}
\caption{Simulated vs. natural exhumation rates: natural samples are represented by the stars. See the legend of Figure 8 for references. Blue points
refer to dry dunite simulations and green points to dry olivine simulations.}
\end{figure}

\begin{figure}
%\begin{landscapefigure}
\noindent\includegraphics[width=28pc]{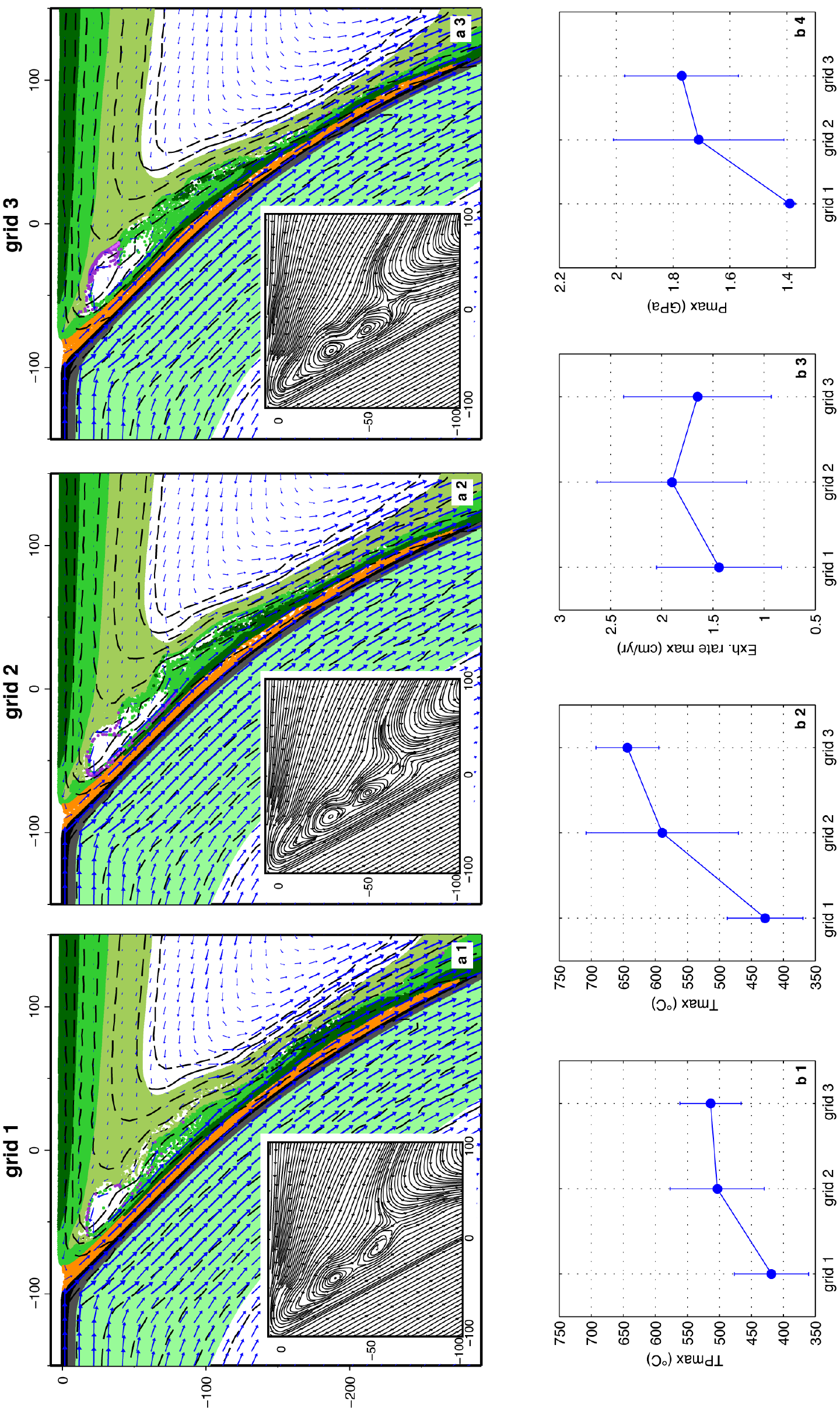}
\caption{Results of the resolution test: in panels $a_i$ the thermodynamic settings for 3 different grids ($a_1$, $a_2$, $a_3$) and the stream lines are displayed; legend as in Figure 2, lithospheric mantle markers are also shown. In panels $b_i$ the statistical parameters, obtained for each grid, are displayed: $T_{Pmax}$, panel $b_1$; $T_{max}$, panel $b_2$; max. exh. rate, panel $b_3$; $P_{max}$, panel $b_4$.}
% \end{landscapefigure}
\end{figure}

\begin{table}[h!]
\caption{Material properties. References: (a) \citep{Ranalli1987}, (b) \citep{Kirby1983}, (c) \citep{Chopra1981}, (d) \citep{Karato1993}, (e)
\citep{Dubois1997,Best2001}, (f) \citep{Rybach1988}, (g) \citep{Gerya2002}.}
\begin{tabular}{lcccccccc}
\hline
\hline
Materials & Rheology &$\mu^0$                             &n& $\rho_0$                & k                         &         $H_r$              & E &
Refs.\\
                 &                   &($Pa\cdot s^{-1}$)    &  &  ($kg/m^3$)           &       (W/mK)        &    ($\mu$W/$m^3$) & (KJ/mol) &\\
\hline
Continental crust & Dry Granite & $3.47\cdot10^{21}$&3.20 & 2640 & 3.01 & 2.50 & 123& a,e,f\\
Oceanic crust &Diabase &$1.61\cdot10^{22}$ &3.40 & 2961 & 2.10 & 0.40 & 260&b,e,f\\
Sediments & $10^{19}$ && & 2640 & 3.01 & 2.50 & 123&e,f\\
Dry mantle & Dry Dunite & $5.01\cdot10^{20}$&3.41 & 3300& 4.15 & 0.002 & 444 &c,e,f\\
		  & Dry Olivine & $5.01\cdot10^{21}$&3.50 & 3300& 4.15 & 0.002 & 540&d,e,f\\
Serpent. mantle & $10^{19}$ & & & 3000 & 4.15 & 0.002 & 444&e,f,g\\
Atmosphere & $10^{23}$       & & & 1.18  &0.026 &            &\\
\hline
\hline
\end{tabular}
\end{table}

\begin{table}
\caption{List of the performed numerical simulations.}
\begin{tabular}{lcccc}
\hline
\hline
Model & Slab velocity & Slab dip  & Dry mantle \\
 ID           &     (cm/y)              &              &      viscosity law \\
\hline
019&1&$30^o$&dry dunite\\
016&3&$30^o$&dry dunite\\
015&5 &$30^o$&dry dunite\\
017&7.5&$30^o$& dry dunite\\
018&10&$30^o$& dry dunite\\
\hline
024a&1&$30^o$& dry olivine\\
021a&3&$30^o$& dry olivine\\
020a&5&$30^o$& dry olivine\\
022a&7.5&$30^o$& dry olivine\\
023a&10&$30^o$& dry olivine\\
\hline
005&1&$45^o$& dry dunite\\
008&3&$45^o$& dry dunite\\
006&5&$45^o$& dry dunite\\
009&7.5&$45^o$& dry dunite\\
010&10&$45^o$& dry dunite\\
\hline
014&1&$45^o$& dry olivine\\
013&3&$45^o$& dry olivine\\
007&5&$45^o$& dry olivine\\
011&7.5&$45^o$& dry olivine\\
012&10&$45^o$& dry olivine\\
\hline
029&1&$60^o$& dry dunite\\
026&3&$60^o$& dry dunite\\
025a&5&$60^o$& dry dunite\\
027&7.5&$60^o$& dry dunite\\
028&10&$60^o$& dry dunite\\
\hline
034&1&$60^o$& dry olivine\\
031&3&$60^o$& dry olivine\\
030&5&$60^o$& dry olivine\\
032a&7.5&$60^o$&dry olivine\\
033&10&$60^o$&dry olivine\\
\hline
\hline
\end{tabular}
\end{table}

\begin{table}
 \caption{Natural cases. References: (1) \citep{Leech2000}, (2) \citep{Ballevre2003}, (3) \citep{Brun2008}, (4) \citep{Rossetti2004}, (5)
 \citep{Thomson1998}, (6) \citep{Ring2003}, (7) \citep{Gosso2010}, (8) \citep{Gazzola2000}, (9) \citep{Thoni1996}, (10) \citep{Patrick1995}, (11)
 \citep{Whitney2008}, (12) \citep{Agard2009}, (13) \citep{Rondenay2008}, (14) \citep{Doglioni2007}, (15) \citep{Zucali2002}.}
 \begin{tabular} {lcccccccccc}
 \hline
 \hline
Location & Unit            & $P_{max}$    & $T_{max}$  & Refs.& Sub.rate& Refs.& Slab      & Refs.& Ex. rate & Refs.  \\
         & Geology         & (GPa)        & ($^oC$)    &     &(cm/y)&     & dip           &      &(cm/y)    &    \\
\hline
Urals    & Unit1            & 1.5-1.7     & 594-637    & (1) &      &      &               &      & 0.07-0.08 & (1)   \\
         & eclog. micasc.   &             &            &     &      &      &               &      &           &     \\
Armorica & Ile de Groix     & 1.8-2.0     & 450 & (2) &  & &  &  &           &     \\
         & micaschists      &             &            &     &      &      &               &      &           &     \\
Calabr.1 & Phyllite and     & 1.2         & $<$400     & (3) & 1-3  & (3)  & $45^¡$-$60^¡$ & (14) & 0.17-0.23 & (3) \\
         & orthogneiss      &             &            &     &      &      &               &      &           &     \\
Calabr.2 & Cont. sediments  & $0.8\pm0.1$ & $400\pm50$ & (4) & 1-3  & (3)  & $45^¡$-$60^¡$ & (14) &0.17-0.23  & (3) \\
         & and marble       &             &            &     &      &      &               &      &           &     \\
Aegean1  & Continental      & $1.0\pm0.2$ & $400\pm50$ & (5) & 1-3  & (5)  &  $30^¡$       & (6)  & 0.4-0.5   & (3) \\
         & phyllite-quartzite &          &            &     &      &      &               &      &           &      \\
Aegean2  & Cyclades         & 1.2-1.9     & 450-550    & (6) & 1-3  & (5)  & $30^¡$        & (6)  & 0.1-0.16  & (6) \\
         & Blueschists      &             &            &     &      &      &               &      &           &     \\
Alps1    & Western Alps     & 1.5-2.2     & 500-600    & (7) & 1-3  & (12) & $30^¡$-$45^¡$ & (14) & 0.14-0.18 & (15)\\
         &Lower Aosta Valley&             &            &     &      &      &               &      &           &     \\
Alps2    & Central Alps,    & 0.9-1.3     & 500-600    & (8) & 1-3  & (12) & $30^¡$-$45^¡$ & (14) &           &     \\
         & Languard-Campo   &             &            &     &      &      &               &      &           &     \\
Alps3    & Koralpe-Saualpe  & 1.7-1.9     & 580-630    & (9) & 1-3  & (12) & $30^¡$-$45^¡$ & (14) &           &     \\
         & Permian Gabbro   &             &            &     &      &      &               &      &           &     \\
Alaska   & Brooks Range     & 0.9-1.2     & 325-415    &(10) & 5    & (13) & $45^¡$        & (13) &           &     \\
         & Orthogneiss      &             &            &     &      &      &               &      &           &     \\
Turkey1  & Menderes Massive & 1.0-1.3     & 450-550    &(11) & 1-3  & (3)  & $30^¡$        & (6)  & 0.08-0.11 &(11) \\
         & metasediment     &             &            &     &      &      &               &      &           &     \\
Turkey2 & Menderes Massive  & 1.0-1.4     & 420-530    &(11) & 1-3  & (3)  & $30^¡$        & (6)  & 0.08-0.11 & (11)\\
        & metasediment      &             &            &     &      &      &               &      &           &     \\
\hline
\hline
 \end{tabular}
 \end{table}

\end{document}